\documentclass[utf8]{frontiersSCNS} 
\usepackage{url,lineno,microtype,subcaption}
\usepackage[onehalfspacing]{setspace}

\usepackage[breaklinks, colorlinks = true,
            linkcolor = SkyBlue,
            urlcolor = SkyBlue,
            citecolor = SkyBlue,
            anchorcolor = white,
            breaklinks]{hyperref}

\def\firstAuthorLast{Oliveira et al.} 
\def\Authors{Denny M. Oliveira\,$^{1,2,*}$, 
			 Mirko Piersanti\,$^{3,4,5}$,
			 Maria-Theresia Walach\,$^{6}$,
			 Livia R. Alves\,$^{7}$,
			 W. Kent Tobiska\,$^{8}$, 
			 Xochitl Blanco-Cano\,$^{9}$, 
			 and Katariina Nykyri\,$^{10}$
			 }
\def\Address{$^{1}$Goddard Planetary Heliophysics Institute, University of Maryland, Baltimore County, 
				Baltimore, MD, United States \\
			 $^{2}$NASA Goddard Space Flight Center, Greenbelt, MD, United States \\
			 $^{3}$Department of Physical and Chemical Sciences, University of L’Aquila, L’Aquila, Italy \\
			 $^{4}$INAF‐IAPS, Rome, Italy \\
			 $^{5}$INFN ‐ Sezione di Roma “Tor Vergata”, via della ricerca scientifica, Rome, Italy \\
			 $^{6}$Lancaster University, Lancaster, LA1 4YW, United Kingdom  \\
			 $^{7}$National Institute of Space Research (INPE), S\~ao Jos\'e dos Campos, Brazil \\
			 $^{8}$Space Environment Technologies, Los Angeles, CA, United States \\
			 $^{9}$National Autonomous University of Mexico, Mexico City, Mexico \\
			 $^{10}$Embry-Riddle Aeronautical University, Daytona Beach, FL, United States
			 }
\def\corrAuthor{Denny Oliveira}

\def\corrEmail{denny.m.deoliveira@nasa.gov}

\begin{document}
\onecolumn
\firstpage{1}

\title[Editorial]{Editorial: Impacts of the Extreme Gannon Geomagnetic Storm of May 2024 throughout the Magnetosphere-Ionosphere-Thermosphere System} 

\author[\firstAuthorLast ]{\Authors} 
\address{} 
\correspondance{} 
\extraAuth{}

\def\By{$B_y$}
\def\Bz{$B_z$}

\def\Ngwira{\href{https://www.frontiersin.org/journals/astronomy-and-space-sciences/articles/10.3389/fspas.2025.1652705/full}{Ngwira et al.}}

\def\Tobiska{\href{https://www.frontiersin.org/journals/astronomy-and-space-sciences/articles/10.3389/fspas.2025.1657731/full}{Tobiska et al.}}

\def\Schennetten{\href{https://www.frontiersin.org/journals/astronomy-and-space-sciences/articles/10.3389/fspas.2024.1498910/full}{Schennetten et al.}}

\def\Lawrence{\href{https://www.frontiersin.org/journals/astronomy-and-space-sciences/articles/10.3389/fspas.2025.1550923/full}{Lawrence et al.}}

\def\DaSilva{\href{https://www.frontiersin.org/journals/astronomy-and-space-sciences/articles/10.3389/fspas.2025.1550635/full}{Da Silva et al.}}

\def\Yuan{\href{https://www.frontiersin.org/journals/astronomy-and-space-sciences/articles/10.3389/fspas.2024.1516222/full}{Yuan et al.}}

\maketitle

{\bf Editorial on the Research Topic} \\
\href{https://www.frontiersin.org/research-topics/65596/impacts-of-the-extreme-gannon-geomagnetic-storm-of-may-2024-throughout-the-magnetosphere-ionosphere-thermosphere-system}{{\bf Impacts of the Extreme Gannon Geomagnetic Storm of May 2024 throughout the Magnetosphere-Ionosphere-Thermosphere System}}

\begin{figure}
		\centering
		\includegraphics[width = 14cm]{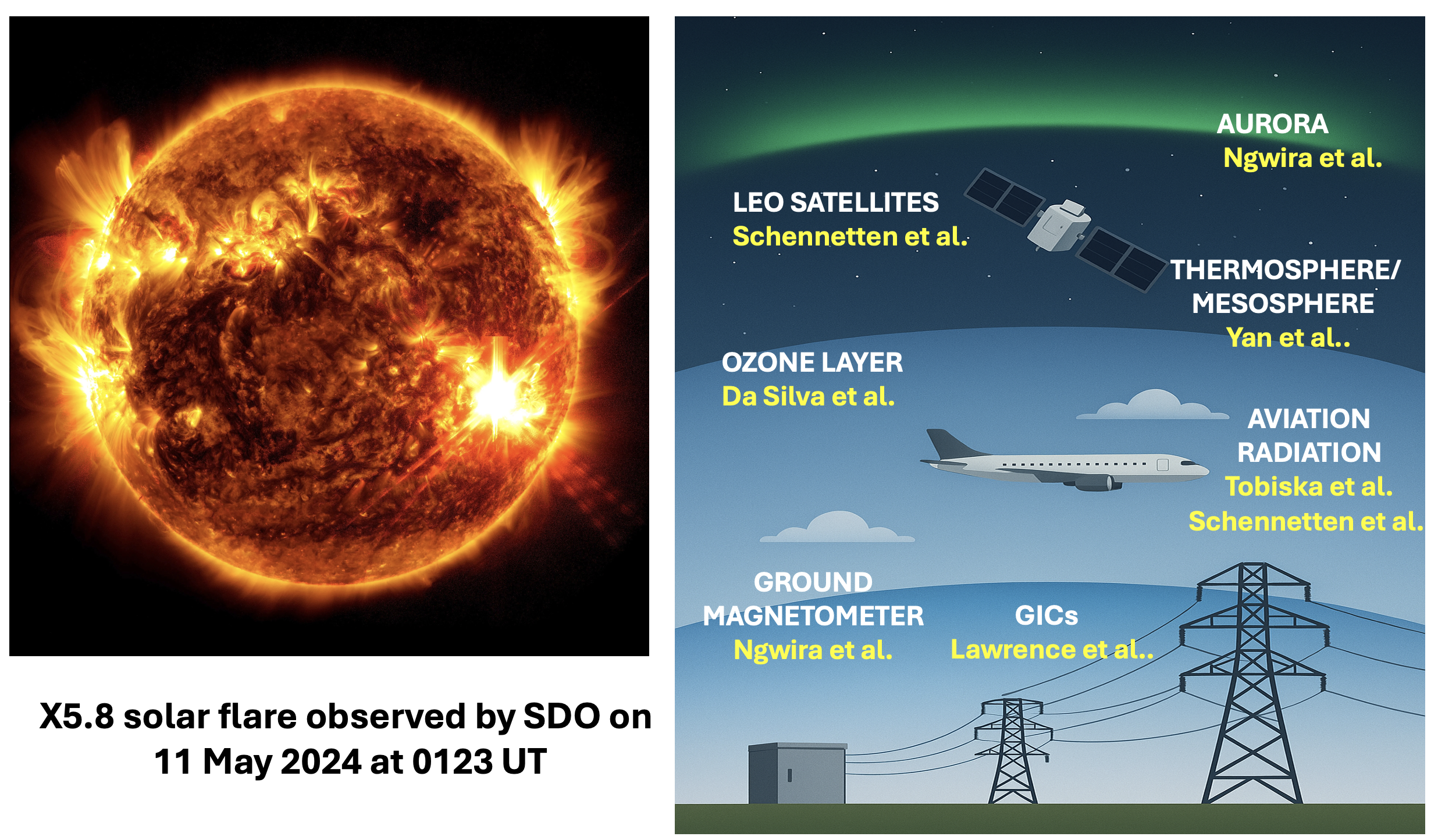}
		\caption{Left: Solar Dynamics Observatory observation of an X5.8 solar flare (bright region near the eastern solar limb a few degrees below the Sun’s equator) associated with Active Region 13664 on 11 May 2024 at 0123 UT. That active region also ejected many CMEs that struck the magnetosphere in early May, inducing the most extreme geomagnetic storms of solar cycle 25. Right: Schematic illustration of key space-weather impacts throughout the geospace environment addressed in this editorial. Effects include thermospheric and mesospheric disturbances, auroral activity, radiation levels measured by satellites at low-Earth orbit, ozone layer variability, aviation radiation exposure, ground magnetic field perturbations, and geomagnetically induced currents in power systems. Text labels highlight the studies addressing each of these domains reported in this editorial.}
		\label{sdo}
	\end{figure}

	\cite{Hayakawa2025} provided a concise data report about the drivers of the extreme geomagnetic storm of May 2024, also known as the Gannon storm. The event was primarily driven by a rapid succession of large coronal mass ejections (CMEs) emanating from a highly active sunspot region, unleashing several X-class solar flares into the interplanetary space. Figure \ref{sdo} (right part), available at \url{https://svs.gsfc.nasa.gov/14703/}, shows a Solar Dynamic Observatory image of the most intense solar flare (X5.8 class) of that period whose peak occurred at 0123 UT of 11 May 2024. The flare is indicated by the intense flash in the image at near the eastern limb slightly below the solar equator. The figure shows a blend of light with wavelengths of 171, 304, and 131 \r{A}, which indicate extreme ultraviolet light. The solar flares and CMEs significantly increased solar energetic particle (SEP) fluxes and radiation levels near Earth, forming temporary radiation belts \citep{Li2025} and elevating high-altitude radiation exposure \citep{Hayakawa2025,Papaioannou2025}. \par

	The multiple CMEs interacted and merged en-route to Earth, resulting in a complex ejecta structure that compressed the magnetosphere (magnetopause standoff position to $\sim$5 Earth radii) and triggered the largest storm in the past two decades (minimum Dst $\sim$ --406 nT). The solar and geomagnetic effects of the May 2024 storm were examined by \cite{TulasiRam2024}, who showed that extremely high solar wind dynamic pressure strongly compressed the magnetosphere. This compression intensified the magnetopause current, the dawn-dusk interplanetary electric field, large sudden impulses, and rapid magnetic field variations observed on the ground \citep{Piersanti2025}. The enhanced pressure also increased solar wind–magnetosphere coupling, strengthening the auroral electrojets and expanding auroral activity to unusually low latitudes \citep{Nilam2025}. Overall, the elevated dynamic pressure was a major factor amplifying the geomagnetic disturbances and ground impacts during the May 2024 storm \citep{TulasiRam2024}. \par

	Many space weather and geomagnetic effects have been reported for this storm. This includes the generation of very large Kelvin-Helmholtz waves at the CME ejecta-sheath boundaries upstream of the Earth \citep{Nykyri2024}, low-latitude auroras \citep{Hayakawa2025}, the largest geomagnetically induced current (GIC) peak ever observed in Latin America \citep{Caraballo2025}, positioning outrages of Global Positioning System (GPS) due to ionospheric disturbances \citep{Yang2025}, caused the largest upward satellite migration ever performed in low-Earth orbit (LEO) \citep{Parker2024a}, and accelerated satellite re-entry from LEO with respect to a reference altitude $\sim$ 280 km \citep{Oliveira2025b}. \par
	
	This Research Topic received 6 articles dealing with different aspects of elevated radiation levels and geomagnetic activity in May 2024. The findings reported in these articles are briefly summarized below.  \par

	We begin this editorial with the topic of aviation radiation effects. \Tobiska{} evaluate how aviation radiation mitigation strategies performed during the extreme Gannon storm due to high solar activity. The authors compare real-time radiation monitoring and mitigation tools at flight altitudes, demonstrating that the established ALARA (as low as reasonably achievable) frameworks successfully limited exposure even as SEP fluxes and atmospheric dose rates surged. Their findings highlight the importance of robust, validated aviation-radiation procedures during extreme space weather events. Such mitigation actions correspond to lowering altitudes and route deviations to equator-ward magnetic latitudes due to magnetic shielding effects \citep{Gao2022}, reinforcing that terrestrial infrastructure and high-altitude aviation are vulnerable -- and yet that mitigation protocols can work effectively under extreme conditions. \par

	Addressing the same topic of aviation radiation, \Schennetten{} used a physics-based radiation transport and dose-calculation model to simulate radiation dose rates at aviation altitudes and LEO, combining three distinct contributions: 1) the additional dose from SEPs; 2) changes in geomagnetic shielding (cut-off rigidities) evaluated during the storm; and 3) the effect of a Forbush decrease reducing the galactic cosmic ray background. Using this approach, \Schennetten{} found that although the SEPs during the Gannon storm did contribute to the radiation field at aviation altitudes, but the dose rates remained relatively low. However, dose rates in LEO reached much higher values. Moreover, when all effects are combined for a hypothetical flight from Frankfurt to Los Angeles, the additional total effective dose was estimated at 14\%–24\% above baseline. The study thereby highlights that while SEPs can elevate doses during an extreme storm, this is not necessarily the case for every such event, because the actual impact depends strongly on the SEP energy spectrum, geomagnetic conditions, and atmospheric shielding. \par

	Using data from the SuperMAG ground network, the Spherical Elementary Current System method -- which maps ionospheric currents -- and mid-latitude auroral imagery, \Ngwira{} examine ground geomagnetic activity during the Gannon storm. Those authors show that powerful CME–CME interactions and prolonged southward interplanetary magnetic field (IMF) B$_\mathrm{z}$ drove exceptionally strong magnetic perturbations -- up to 700 nT and $>$500 nT/min at mid-latitudes across the United States. The auroral oval expanded equatorward to ~40$^\circ$–45$^\circ$ magnetic latitude, producing vivid aurorae and intense substorm activity far from the polar regions in the American sector \citep{Gonzalez-Esparza2024,Hayakawa2025}. The results reveal how extreme IMF conditions and abrupt B$_\mathrm{y}$ rotations can trigger westward-propagating auroral electrojets that enhance ground-level magnetic variations, underscoring the vulnerability of mid-latitude infrastructure to space weather hazards once thought confined to high latitudes. \par

	\Lawrence{} present both direct ground magnetic and electric field observations across the UK and combine these with magnetotelluric-derived resistivity models of the subsurface and a high-voltage transmission network model to estimate GICs. The key findings include the auroral electrojet intrusion to surprisingly low latitudes (below $\sim$54$^\circ$N) across central and southern England, rapid and large ground magnetic field changes (minute-to-minute magnetic field variations) and estimated substation GICs exceeding $\sim$60 A in parts of southwest and east-central England and northern Wales. As a result, \Lawrence{} show that the integration of multi-site geoelectric modeling, grid-network simulation and magnetic observations offers a comprehensive assessment of ground-impact risks during extreme space weather and highlights the need for region-specific vulnerability analyses. \par

	\DaSilva{} combine multi-instrument satellite and ground-based observations to investigate how the Gannon storm affected the South Atlantic Magnetic Anomaly (SAMA), a region with the weakest geomagnetic field \citep{Gledhill1976}. Using particle detectors aboard LEO satellites, \DaSilva{} monitored the behavior of low-energy electrons in the inner radiation belt, while ionospheric radars and digisondes provided measurements of ionization changes over South America. In addition, satellite-borne sensors tracked variations in ozone concentration in the upper atmosphere. The study found that the flux of low-energy electrons in the inner radiation belt vanished at the same time that energy input into the magnetosphere ceased -- indicating a strong coupling between these processes. This period coincided with enhanced ionization in the ionospheric E-region over SAMA and pronounced variability in electron fluxes. \par 

	Finally, \Yuan{} used coordinated lidar and optical observations from a midlatitude station to reveal how the neutral atmosphere responded to the Gannon storm. Their analysis show that during the storm, the upper portion of the sodium layer nearly disappeared at the same time that thermospheric temperatures increased and strong equatorward winds developed. These simultaneous changes indicate that energy and momentum from high-latitude geomagnetic activity penetrated deeply into midlatitudes, directly altering both the dynamics and composition of the mesosphere and lower thermosphere. The results demonstrate that even regions far from the auroral zones can experience significant coupling between magnetospheric forcing, neutral winds, and atmospheric chemistry during extreme space weather events. The right panel of Figure \ref{sdo} presents a schematic diagram highlighting the key regions and technological systems affected by space weather, along with the first author associated with each contributing study. \par

	In summary, this Research Topic enhances our understanding of solar and geomagnetic effects during the May 2024 storm on aviation radiation, GICs and geoelectric fields in the US and the UK, effects on the atmosphere at mid-latitude mesosphere and lower thermosphere, and perturbations in the neutral and ionized atmosphere over the SAMA region. Although our focus was on the Gannon storm, the October 2024 storm can also offer further insights of extreme space weather effects. Even though the October event was the third extreme event since the Halloween storms of 2003 \citep{Oliveira2025a}, very few studies have focused on that event thus far. Examples are intense red aurora observations by citizen scientists \citep{Kataoka2025}, impacts of merging electric field on field-aligned currents and polar electrojets \citep{Xia2025}, and a premature re-entry of a Starlink satellite \citep{Oliveira2025a}.  Studies comparing many aspects of the extreme 2024 events -- including their drivers and subsequent space weather effects --  could be the focus of a future Rearch Topic in {\it Frontiers in Astronomy and Space Sciences}. \par


	Jennifer L. Gannon was a distinguished space-weather scientist whose research profoundly advanced our understanding of GICs and ground-magnetic disturbances \citep{Lugaz2024}. Dr. Gannon's work established essential connections between magnetospheric physics and the protection of technological infrastructure on Earth. The extreme geomagnetic storm of May 2024, now known as the Gannon storm and the topic of this editorial, fittingly honors her enduring contributions and impact on the geospace community. We dedicate this editorial and Research Topic to the memory of Dr. Gannon.

\section*{Author Contributions}

	All authors listed in this manuscript have made a substantial, direct, and intellectual contribution to the work and approved it for publication. All co-authors proofread and approved this manuscript.

\section*{Funding}
	
	DMO acknoledges funding provided by UMBC's START (Strategic Awards for Research Transitions) program (grant code SR25OLIV), NASA's Living With a Star (LWS) program (NNH22ZDA001N-LWS), NASA's Internal Scientist Funding Model (ISFM), and NSF's EMpowering BRoader Academic Capacity and Education (EMBRACE) program (AGS-2432908). MTW acknowledges funding from UKRI through the ErnestRutherford Fellowship (ST/X003663/1). LRA acknowledges funding from CNPq grant PQ-C- 304482/2024-2. KN acknowledges funding from NASA LWS award $\#$80NSSC23K0899 and NASA ISFM, and Magnetosphere Multiscale (MMS) mission programs.



\section*{Conflict of Interest Statement}

	The authors declare that the research was conducted in the absence of any commercial or financial relationships that could be construed as a potential conflict of interest.

\section*{Publisher's note}
	
	All claims expressed in this article are solely those of the authors and do not necessarily represent those of their affiliated organizations, or those of the publisher, the editors and the reviewers. Any product that may be evaluated in this article, or claim that may be made by its manufacturer, is not guaranteed or endorsed by the publisher.

\bibliographystyle{frontiersinSCNS_ENG_HUMS} 

\end{document}